\documentclass[pre, twocolumn, showkeys, amsmath, amssymb]{revtex4}

\usepackage{graphicx}

\begin{document}

\title{A hypothesis on the selective advantage for sleep}

\author{Emmanuel Tannenbaum}
\email{emanuelt@bgu.ac.il}
\affiliation{Department of Chemistry, Ben-Gurion University of the Negev,
Be'er-Sheva 84105, Israel}

\begin{abstract}

In this note, we present a hypothesis for the emergence of the phenomenon of
sleep in organisms with sufficiently developed central nervous systems.
We argue that sleep emerges because individual neurons must periodically
enter a resting state and perform various ``garbage collection'' activities.
Because the proper functioning of the central nervous system is dependent
on the interconnections amongst a large collection of individual neurons,
it becomes optimal, from the standpoint of the organism, for these
garbage collection activities to occur simultaneously.  We present
analogies with complex structures in modern economies to make our case,
although we emphasize that our hypothesis is purely speculative at
this time.  Although the ``garbage collection'' hypothesis
has been previously advanced, we believe that our speculation is useful 
because it illustrates the power of a general paradigm for understanding the emergence
of collective behavior in agent-built systems.

\end{abstract}


\maketitle

Sleep is one of the most intriguing phenomena exhibited by multicellular
organisms.  It occurs in almost all organisms with a central nervous 
system, and is characterized by a period during which there is a marked decrease
in the arousal level of the organism.                                                     

The difficulty in understanding the necessity of sleep derives from the
observation that sleep-deprived organisms do not show signs of physical
damage, and yet they are characterized by impaired cognitive abilities.
In extreme cases, sleep deprivation can even lead to death.

A hypothesis for sleep, advanced by Francis Crick, is that sleep is
a way for the brain to perform various upkeep, or alternatively,
``garbage collecting'' functions necessary for proper brain function
\cite{SLEEP}.  More specifically, sleep is a time when the brain sorts 
through various stored memories, and discards those deemed unessential, while 
processing those deemed essential for long-term storage.

There is now evidence suggesting that Crick's hypothesis may be correct:
It has been discovered that neurons contain a protein, termed Fos,
which is involved in proper neuronal function \cite{SLEEP}.  During periods of
neuronal stimulation, Fos naturally builds up, apparently as a by-product
of various neuronal activities.  Proper neuronal function is no longer
possible once Fos levels reach a critical level.  During the sleep state of 
an organism, Fos levels rapidly drop.  Apparently, the Fos protein acts as 
a molecular switch that regulates various genes involved in proper neuronal 
function \cite{SLEEP}.  During sleep, these genes are suppressed, allowing
the neuron to re-set for the next period of wakefulness.

We should note a parallel between this behavior and the build-up of lactic
acid during intense muscular activity.  The accumulation of lactic acid
causes the muscles to tire, and to cease functioning, before permanent damage
occurs.  Similarly, it is apparent that Fos itself may not be necessary
for the survival of the neuron, but rather acts as a chemical signal that
regulates external input to the neuron in such a way to prevent neuronal
damage.

While there is emerging evidence that individual neurons may
periodically require periods of limited external stimulation
in order to ``reset'' themselves, this in and of itself
does not explain sleep as collective synchronization phenomenon, 
whereby the neurons of the central nervous systems engage in
their respective upkeep tasks simultaneously.

Here, we argue that the drive for such ``swarming'' behavior is that 
it is optimal, from the standpoint of brain function and therefore
organismal survival, for neurons to engage in their
respective ``garbage collection'' activities together.  This optimal
garbage collection strategy is derived from the highly connected,
interdependent nature of neurons in a central nervous system.  
Therefore, if some neurons were to engage in garbage collection
independently of others, this would impede the proper functioning
of other neurons, who may rely on the ``sleeping'' neurons for
important information. 

To understand this argument in terms of a more general paradigm,
we note a recent speculative paper by the author regarding the
nature of RNA biochemical networks in eukaryotic cells
\cite{RNACOMMUNITY}.  In this work, the author argued that
the RNA biochemical networks in eukaryotic cells may in 
many ways be viewed as an RNA-''community'', in the sense
that agent-based models will likely be as useful in understanding
the structure of such biochemical networks as the traditional
pathway-based approach of Systems Biology.  Because the emergence
of other complex structures, such as the brain and highly networked
societies, is driven by the self-organization of agents acting under 
certain selection pressures, it was then argued that the emergence
of complexity may be seen as the outcome of a series of agent-based
behavior at various length scales, where the collective interactions
of agents at one length scale produce a single agent at the next 
length scale.  Thus, there are likely analogous structures and
behaviors amongst the various length scales, so that, by observing
such structures at one length scale, it may be possible to infer
or understand similar structures at another length scale.
 
To apply this to the problem of sleep, we use two examples, that
of a city, and that of a large corporation.  First of all, we note
that, just as with complex organisms, a city has a periodic cycle of
activity, characterized by high levels of activity during the day,
followed by periods of relatively low levels of activity at night.
The central reason for this collective behavior is that the individual
agents constituting a city, namely humans, are awake during the day
and sleep at night.  Since the brain is a collection of neurons
whose interconnections are driven by a biochemical reward-punishment
system, it suggests that the brain requires sleep because individual
neurons require periods of reduced activity.

An analogy with a large corporpation is even more
useful for gaining insight.  A large corporation is often
divided into numerous branches (even a smaller company may have 
many distinct departments), all of which are required to interact
with one another to a certain extent, in order for the corporation
to function properly.  Focusing on one such subdivision, it is natural
to suppose that during the normal operations of the subdivision, 
various transactions occur which are temporarily recorded using
locally available materials (keeping a running receipt of the day's
purchases, for example).  However, for proper functioning
of the subdivision, and of the corporation in general, which requires
up-to-date, consolidated information from its various components, 
it makes sense for each subdivision to periodically sort through
its various transactions and update necessary information, such
as inventories and financies, for later use.  Once this update
has been performed, much of the information from the numerous
smaller transactions can be discarded.

If the various subdivisions of a corporation are highly interconnected,
it will likely be most efficient if the various subdivisions engage
in their various ``housekeeping'' tasks simultaneously.  First of all,
this guarantees that all the subdivisions of a corporation will be operational
at the same time.  Secondly, if the various subdivisions of a corporation do not perform their upkeep tasks
simultaneously, then one subdivision might make use of information
from another subdivision before the information has been properly
updated, and the result can be a cascade of temporally out-of-phase
information transfer that can cause a system failure.  While this
may not manifest itself as physical damage to the actual structures
of the corporation, it may lead to an inability for the corporation
to properly interact with the external environment, causing it to 
fail.

The applicability of this analogy to the brain should be clear:
Because the brain is a highly interconnected computational network
of neurons, for the brain to properly process external information
it is necessary that the information stored in the various parts
of the brain be temporally in-phase.  Furthermore, during periods
of wakefulness, the brain will function optimally if all neurons
are fully engaged in brain activity.

Our analysis is not complete, however, since, due
to modern technology, there are now corporations that
can perform their upkeep tasks in real time, without having to
shut-down.  Essentially, modern technology has reached a point
where information transfer within a corporation is faster than
the rate of generation of information due to external inputs, so 
that periodic shut-down and performance of upkeep tasks are not 
necessary (however, even in this case, there are likely other 
time scales on which shut-down may be necessary).

To resolve this apparent contradiction for our explanation for
sleep, we first point out the possibility that, at least in some organisms that sleep,
neuronal biochemistry and brain circuitry will evolve to
a point where only minimal sleep will be required for
proper organismal function.

Second, while it is in principle possible for 
individual neurons to continuously switch between various
information transfer, processing, and consolidation functions,
it is energetically costly to do so, since the neuron would have
to maintain the various pathways turned on at all times.  Furthermore,
the switching strategy itself has an energetic penalty.  

Therefore, sleep is also an example of the general idea that when a system
is faced with a requirement to continuously perform a given set of tasks,
it is often the least energetically costly strategy to
complete each of the tasks sequentially, instead of continually
switching from one to the other.  The optimality of this strategy
increases as the base rate at which tasks must be performed 
approaches the saturation point of the system.  That natural
selection drove central nervous systems to process information
close to their saturation point is not unexpected, since
for a given energy input, those organisms that can process
more information will likely have a greater survival advantage.

We conclude with an additional hypothesis that the synchronized sleep behavior
in neurons is explicitly encoded into neural genomes.  By analogy with
cancer and other phenomena involving defection from collective behavior,
we also hypothesize that there exist sleep disorders where some neurons,
either through defective regulatory genes or some other external influence,
do not coordinate their sleep behavior with other neurons, and engage in
their upkeep tasks independently of the rest of the brain.  Just as
with tumor cells, there are likely always a few neurons who engage in this behavior.
However, if this behavior becomes widespread, with the brain divided into 
numerous distinct regions, each of which ``sleep'' on their own, the result
could be a sleep disorder that is potentially life-threatening.  We also claim
that this hypothesized ability of individual neurons to ``sleep'' independently
of the rest of the brain may occur as a neuronal stress response, triggered
by overstimulation of the central nervous system.

\begin{acknowledgments}

This research was supported by the Israel Science Foundation.  The author would like to thank 
Allen Tannenbaum (Georgia Institute of Technology and Technion -- Israel Institute of Technology) 
for helpful conversations regarding this work.

\end{acknowledgments}

\end{document}